\documentclass[preprint,showpacs,aps]{revtex4}
\usepackage{graphicx}
\usepackage{epsfig}
\usepackage{verbatim}
\usepackage{dcolumn}
\usepackage{amsmath}

\begin{document}
\title{New Relations for Coefficients of Fractional Parentage---the Redmond
Recursion Formula with Seniority}

\author{L. Zamick and A. Escuderos}

\affiliation{Department of Physics and Astronomy, Rutgers University,
Piscataway, New Jersey USA 08854}

\date{\today}

\begin{abstract}
We find a relationship between coefficients of fractional parentage (cfp)
obtained on the one hand from the principal parent method and on the other hand
from  a seniority classification. We apply this to the Redmond recursion 
formula which relates $n \to n+1$ cfp's to $n-1 \to n$ cfp's where the 
principal-parent classification is used. We transform this to the seniority 
scheme. Our formula differs from the Redmond formula inasmuch as we have a sum 
over the possible seniorities for the $n \to n+1$ cfp's, whereas Redmond has 
only one term. We show that there are useful applications of both the principal
parent and the seniority classification.
\end{abstract}
\pacs{}
\maketitle

\section{Introduction}

The wave function of a system of identical particles is antisymmetric in space
and spin. It is, however, often convenient to single out a given particle. This
can be done by means of a fractional parentage expansion. For $n$ identical
particles in a single $j$ shell, the expansion is as follows
\begin{equation}
\Psi (1 \cdots n)^{I \alpha} = \sum_{J_1 \alpha_1} [j^{n-1} J_1 \alpha_1 j
|\} j^n I \alpha ] \left\{ \Psi (1 \cdots n-1 )^{J_1 \alpha_1} \cdot \Psi (n)^j
\right\}^{I \alpha} , \label{npart}
\end{equation}
where the curly brackets designate a Clebsch--Gordan coupling and the 
quantities in square brackets are the coefficients of fractional parentage. 
Although the total wave function  is antisymmetric, each term in the expansion
is not. A given term is antisymmetric in the first $n-1$ particles, but not in
the $n$-th. In some sense, with cfp's we have our cake and we eat it too.

Such an expansion enables one to single out a certain particle despite the fact
that the total wave function is antisymmetric. A simple example for the use of 
a cfp involves transfer reactions. The cross section for the pickup of a 
neutron from a shell which has $n$ neutrons and no protons is proportional to a
spectroscopic factor, the value of which is
\begin{equation}
{\cal S} = n [ j^{n-1} (J_f \alpha_f)~j |\} j^n J_i \alpha_i ]^2 ~,
\end{equation}
where $(J_i~\alpha_i)$ refer to the $(n+1)$-neutron system and $(J_f~\alpha_f)$
to the $n$-neutron system. The summed pickup strength over all $(J_f~\alpha_f)$
is equal to $n$, the number of particles available to be picked up.

One method of calculating coefficients of fractional parentage is by the
principal-parent technique. An explicit example for a system of three identical
particles is given by de~Shalit and Talmi~\cite{st63} (see Eq.~26.11 on page
271) and will be repeated in this introduction in order to establish notation.

To get a cfp for three identical particles, one first combines two of them to
a total angular momentum $J_0$: $[j j]^{J_0}$. We call this the principal 
parent. We then add a third particle and, after antisymmetrizing and 
normalizing, the resulting wave function is
\begin{equation}
\Psi^J [J_0] = N [J_0] (1-P_{12}-P_{13}) \left[ [12]^{J_0} 3 \right]^J,
\label{antiwf}
\end{equation}
where $N [J_0]$ is the normalization factor. We then expand~\eqref{antiwf} as
per Eq.~\eqref{npart}, obtaining
\begin{equation}
[j^2(J_1) j J |\} j^3 [J_0]J] = N [J_0] \left[ \delta_{J_0 J_1} + 2
\sqrt{(2J_0+1) (2J_1+1)} \begin{Bmatrix} j & j & J_0 \\ J & j & J_1 
\end{Bmatrix} \right] , \label{ppcfp23}
\end{equation}
where
\begin{equation}
N [J_0] = \left[3+6(2J_0+1) \begin{Bmatrix} j & j & J_0 \\ J & j & J_0
\end{Bmatrix} \right]^{-1/2}. \label{norm}
\end{equation}
Note the relationship between the cfp and the normalization factor:
\begin{equation}
[j^2(J_0) j J |\} j^3 [J_0] J ] = \frac{1}{3 N [J_0]}. \label{cfpnf}
\end{equation}

A recursion formula for cfp's due to Redmond~\cite{r54} is presented in the 
books of Talmi and de~Shalit~\cite{st63} on page 528, and Talmi~\cite{t93}
on page 274. It can be written as follows

\begin{eqnarray}
\lefteqn{(n+1) [j^n (\alpha_0 J_0) j J |\} j^{n+1} [\alpha_0 J_0] J]
[j^n (\alpha_1 J_1) j J |\} j^{n+1} [\alpha_0 J_0] J ] =} \nonumber \\
 & = & \delta_{\alpha_1 \alpha_0} \delta_{J_1 J_0} +
n (-1)^{J_0 + J_1} \sqrt{(2J_0+1)(2J_1+1)} \sum_{\alpha_2 J_2}
\begin{Bmatrix} J_2 & j & J_1 \\ J & j & J_0 \end{Bmatrix} \nonumber \\
 & & \times [j^{n-1} (\alpha_2 J_2) j J_0 |\} j^n \alpha_0 J_0] 
[j^{n-1} (\alpha_2 J_2) j J_1 |\} j^n \alpha_1 J_1]. \label{redmond}
\end{eqnarray}

In the above, square bracket designates the principal parent used to calculate
the cfp. Actually, the principal parent sometimes looses its significance
because in some cases more than one principal parent can yield the same cfp.
In tables of cfp's, the principal parent is usually not listed. The quantities
in parentheses $(\alpha_0 J_0)$ are listed. The cfp with $(\alpha_0 J_0)$ is
the probability amplitude that a system of $(n+1)$ identical particles can be
separated into a system of $n$ particles with quantum numbers $(\alpha_0 J_0)$
and a single nucleon.

Note that the $n-1 \to n$ cfp's on the right-hand side of Eq.~\eqref{redmond}
do not have principal-parent quantum numbers. Indeed they are not fully 
specified, i.e., the quantum number $\alpha$ is not defined. Althought not
necessary, we shall assume $\alpha$ stands in part for the seniority quantum
number and that the $n-1 \to n$ cfp's form a complete orthonormal set.

One problem with the principal-parent method is that one gets more cfp's than
there really are. One can see this, for example, from Eq.~(\ref{ppcfp23}). For
$j=7/2$ in $^{43}$Ca, the allowed states have total angular momentum $I=3/2,
5/2,7/2,9/2,11/2$, and $15/2$, all occurring only once. Consequently, there is 
only one set of cfp's for each angular momentum. If we construct the cfp's for
the unique $I=7/2$ state of the $f_{7/2}^3$ configuration, using first 
$J_0=2$ and then $J_0=4$ as principal parents, we get exactly the same cfp's 
for $J_1=0,2,4$, and $6$ in the two cases. There is a redundancy. 

For the $j^3$ configuration with $j=9/2$, $I=9/2$, there are two states. In
the Bayman--Lande scheme~\cite{bl66}, the states are classified by the 
seniority quantum number. Of the two $I=9/2$ states above, one has seniority 1
and the other has seniority 3. However, in the principal parent scheme, there
are five sets of cfp's corresponding to $J_0=0,2,4,6$, and $8$. This is clearly
an overcomplete set.

In Table~\ref{tab:pps-bl} we show the results of the two schemes for the 
example above: $I=j=9/2$.

\begin{table}[ht]
\caption{Coefficients of fractional parentage in the principal-parent scheme
and in the seniority scheme (results from Bayman--Lande) for the $I=9/2$ states
of the $g^3_{9/2}$ configuration.} \label{tab:pps-bl}
\hfill
\begin{tabular*}{.5\textwidth}[t]{@{\extracolsep{\fill}}lrclr}
\toprule
\multicolumn{5}{c}{Principal-parent scheme} \\ \colrule
\multicolumn{2}{c}{$J_0=0$} \\
$J_1=0$ & 0.516398 \\
$J_1=2$ & $-0.288675$ \\
$J_1=4$ & $-0.387298$ \\
$J_1=6$ & $-0.465475$ \\
$J_1=8$ & $-0.532291$ \\ 
\multicolumn{2}{c}{$J_0=2$} & & \multicolumn{2}{c}{$J_0=4$} \\
$J_1=0$ & $-0.437384$ & & $J_1=0$ & $-0.262994$ \\
$J_1=2$ & 0.340825 & & $J_1=2$ & $-0.008910$ \\
$J_1=4$ & $-0.019881$ & & $J_1=4$ & 0.760475 \\
$J_1=6$ & 0.764610 & & $J_1=6$ & $-0.362496$ \\
$J_1=8$ & 0.327887 & & $J_1=8$ & 0.470138 \\
\multicolumn{2}{c}{$J_0=6$} & & \multicolumn{2}{c}{$J_0=8$} \\
$J_1=0$ & $-0.286884$ & & $J_1=0$ & $-0.473618$ \\
$J_1=2$ & 0.311026 & & $J_1=2$ & 0.192553 \\
$J_1=4$ & $-0.329013$ & & $J_1=4$ & 0.616034 \\
$J_1=6$ & 0.837866 & & $J_1=6$ & 0.149270 \\
$J_1=8$ & 0.103396 & & $J_1=8$ & 0.580370 \\
\botrule
\end{tabular*}
\hfill
\begin{tabular*}{.22\textwidth}[t]{@{\extracolsep{\fill}}lr}
\toprule
\multicolumn{2}{c}{Seniority scheme} \\ \colrule
\multicolumn{2}{c}{$v=1$} \\
$J_1=0$ & $-0.516398$ \\
$J_1=2$ & 0.288675 \\
$J_1=4$ & 0.387298 \\
$J_1=6$ & 0.465475 \\
$J_1=8$ & 0.532291 \\ \\
\multicolumn{2}{c}{$v=3$} \\
$J_1=0$ & 0.000000 \\
$J_1=2$ & 0.181186 \\
$J_1=4$ & $-0.654463$ \\
$J_1=6$ & 0.696673 \\
$J_1=8$ & $-0.231293$ \\
\botrule
\end{tabular*} \hspace*\fill
\end{table}

Of course, one can have more than one state of a given seniority. For example,
for $j=15/2$, $(j^3)^{I=15/2}$, there is one state of seniority 1 and two of
seniority 3.

The discussion of seniority (a topic introduced by G.~Racah \cite{r43,r49}) is
given extensively in text books \cite{st63,t93,l80}, so we will be very brief
on this. To simplify the discussion, let us consider a closed shell of protons
and focus only on the open-shell system of neutrons, i.e., deal only with
identical particles. For an even number of neutrons, there is
a tendency for their spins to be paired. This corresponds to a
seniority $v=0$ state with total angular momentum 0 (note that all even--even
nuclei have angular momentum 0). To form a $2^+$ state, one must break at 
least one pair. As noted by Lawson, the seniority $v$ of a nuclear state ``is 
the number of unpaired nucleons in the eigenfunction describing the 
state''~\cite{l80}. He also mentions that the delta-function potential 
conserves seniority in a single 
$j$ shell. For an even number of neutrons, the seniority $v$ must be an even 
integer; for an odd number of neutrons, it must be an odd integer. In the case
of a semimagic nucleus with
an open shell of, say, neutrons, whereas the $I=0$ ground state has dominantly
$v=0$, the first $2^+$ state is dominantly $v=2$. However, for the $I=4$ state
in $^{44}$Ca ($f_{7/2}^4$), the seniority $v=4$ state is slightly lower than
the seniority $v=2$ state. One can understand this by noting that the $v=2$ 
state consists of one broken pair with $J=4$, while the seniority $v=4$ state
can be constructed from two $J=2$ pairs. For a two-particle system, the $J=2$
pair energy is lower than the $J=4$ pair energy often by a factor of two or
more.

\section{Relation between principal parent cfp's and those in the seniority
scheme}

We here note a relationship between the overcomplete set of principal-parent 
coefficients of fractional parentage and those with the seniority 
classification:

\begin{eqnarray}
\lefteqn{[j^n (v_0 J_0) j J |\} j^{n+1} [v_0 J_0] J]
[j^n (v_1 J_1) j J |\} j^{n+1} [v_0 J_0] J] =} \nonumber \\
 & = & \sum_v [j^n (v_0 J_0) j J |\} j^{n+1} J v ]
[j^n (v_1 J_1) j J |\} j^{n+1} J v ] . \label{pp2scfp}
\end{eqnarray}
In the left-hand side above, the first principal parent is formed by
adding the ($n+1$)-th nucleon to an $n$-nucleon antisymmetric system with good 
seniority and angular momentum ($v_0$ $J_0$), then coupling the combined system
to a total angular momentum $J$, and then antisymmetrizing and normalizing the 
total wave function. On the right-hand side, the sum over $v$ is a sum over all
the possible seniorities of the combined ($n+1$) system and, for a given 
seniority, over all states with that seniority.

A proof of the above result will be given in Appendix~\ref{app}.

We can verify the result of Eq.~\eqref{pp2scfp} for specific examples. Consider
first a system of three identical particles in a $j=15/2$ shell with total 
angular momentum $J=15/2$. Take the principal-parent angular momentum $J_0$ to 
be equal to 2, and also $J_1=2$. Using the explicit formulae of 
Eqs.~\eqref{ppcfp23} and \eqref{norm}, we obtain
\begin{equation}
[j^2(J_0) j J |\} j^3[J_0]J]^2=\frac{1}{3} \left[ 1+2(2J_0+1)
\begin{Bmatrix} j & j & J_0 \\ J & j & J_0 \end{Bmatrix} \right] = 0.153945.
\end{equation}

From Bayman and Lande~\cite{bl66} we find
\begin{eqnarray}
[j^2 (2) j J=15/2 |\} j^3 v=1] & = & 0.172516 , \\
{[}j^2 (2) j J=15/2 |\} j^3 v=3, \alpha=1{]} & = & 0.153452 , \\
{[}j^2 (2) j J=15/2 |\} j^3 v=3, \alpha=2{]} & = & 0.317231 .
\end{eqnarray}
We easily verify that the sum of the squares is 0.153945.

As a second example, consider the case $j=9/2$, $J=9/2$, with $J_0=[J_0]=2$ 
and $J_1=4$. The left-hand side of Eq.~(\ref{pp2scfp}) is given by
\begin{equation}
\text{lhs}=\frac{4}{3} \sqrt{45}
\begin{Bmatrix}
9/2 & 9/2 & 2 \\ 
9/2 & 9/2 & 4
\end{Bmatrix} = -0.00677596 .
\end{equation}
The right-hand side has contributions from $v=1$ and $v=3$. Using the 
Bayman--Lande tables, we find
\begin{subequations}
\begin{eqnarray}
v=1 & & 0.288675 \times 0.387598 \\
v=3 & & -0.181166 \times 0.654463 \\
\text{Total} & & -0.006776 
\end{eqnarray}
\end{subequations}

\section{An example of the use of the overcomplete cfp's}

Ironically, one can get the most useful information from principal parent
cfp's by calculating them for states which do not exist. We use 
Eq.~(\ref{ppcfp23}) to illustrate this point. As noted by Racah~\cite{r43,r49},
de Shalit and Talmi~\cite{st63} and Talmi~\cite{t93}, there are no states of 
the $j^3$ configuration with total angular momentum $J=3j-4$. If in 
Eq.~(\ref{ppcfp23}) we choose the principal parent $J_0=2j-1$ and $J_1=2j-3$,
then the fact that the cfp does not exist leads to the relation
\begin{equation}
\begin{Bmatrix}
j & j & (2j-1) \\
(3j-4) & j & (2j-3)
\end{Bmatrix} =0~.
\end{equation}
For a different choice, $J_0=J_1$, one gets
\begin{equation}
\begin{Bmatrix}
j & j & J_0 \\ J & j & J_0
\end{Bmatrix} = - \frac{1}{2(2 J_0+1)} \label{j0eqj1}
\end{equation}
for certain states $J$ that do not exist in the $j^3$ configuration. Note that,
for these select $J$ values, this 6$j$ symbol does not depend on what $J$ is.
For example, for $j=7/2$, Eq.~\eqref{j0eqj1} holds for $J=1/2,13/2,17/2$, and 
19/2, but not for the allowed $(f_{7/2}^3)$ states mentioned previously.

An interesting use of these 6$j$-symbol relations has been found by Robinson
and Zamick~\cite{rz} for a system of two neutrons and one proton (or two 
protons and one neutron), e.g., $^{43}$Sc ($^{43}$Ti) for $j=7/2$. To perform
a shell model calculation, one uses as input two-body matrix elements $\langle
(j_1 j_2)^{JT} | V | (j_3 j_4)^{JT} \rangle$, where $J$ is the total 
two-particle angular momentum and $T$ is the isospin. Of course, $T$ can only
be either zero or one for a two-particle system. The resulting wave function 
for $^{43}$Sc in the single $j$ shell can be written as
\begin{equation}
\Psi^I = \sum_{J_N} D^I(j_\pi,J_N) ~ [j_\pi (j^2)^{J_N} ]^I~,
\end{equation}
where, for a state of total angular momentum $I$, $D^I(j_\pi,J_N)$ is the 
probability amplitude that the neutrons couple to $J_N$ ($J_N=0,2,4$, or 6).

Without going into detail (these are given in Ref.~\cite{rz}), the authors
considered a model in which the two-body matrix elements with isospin $T=0$
were set equal to zero. Only the $T=1$ two-body matrix elements entered into
the calculation. When this was done, an interesting partial dynamical symmetry 
was found for the previously mentioned angular momenta $I$ which cannot occur
for a $j^3$ configuration of identical particles, namely $I=1/2,13/2,17/2$, 
and 19/2. It was found for these states that $J_N$ was a good quantum number 
for the wave functions, i.e., a given state wave function was of the form
$[j_\pi~J_N]^I$. As an example, for $I=13/2$ the matrix element $\langle [j~4
]^{13/2} | V | [j~6]^{13/2} \rangle$ was zero. This is explained by the 
vanishing of the 6$j$ symbol of Eq.~\eqref{norm}
\begin{equation}
\begin{Bmatrix}
7/2 & 7/2 & 6 \\ 3/2 & 7/2 & 4
\end{Bmatrix} =0 ~,
\end{equation}
which we remember was obtained by completely different considerations.

There were also degenerate states, such as $I=1/2^-$ and $13/2^-_1$, whose
wave functions were of the form $[j_\pi=7/2, J_N=4]^I$. Likewise, $13/2^-_2, 
17/2^-$, and $19/2^-$ were all degenerate with wave functions $[j_\pi=7/2,
J_N=6]^I$. These degeneracies follow from Eq.~\eqref{j0eqj1}.

We call the above a {\it partial} dynamical symmetry because it applies only
to states of angular momentum $I$ which {\it can} occur for a system of two
neutrons and  a proton, but cannot occur for a system of three neutrons (or
three protons).

\section{The Redmond recursion relation in the seniority scheme}

We here present the equivalent of the Redmond recursion relation, but for
cfp's classified by the seniority quantum number $v$ and for which there are no
redundacies. Here is our formula

\begin{eqnarray}
\lefteqn{(n+1) \sum_{v_s} [j^n (v_0 J_0) j I_s|\} j^{n+1} v_s I_s]
[j^n (v_1 J_1) j I_s |\} j^{n+1} v_s I_s] =} \nonumber \\
 & = & \delta_{J_0 J_1} \delta_{v_0 v_1} 
+ n (-1)^{J_0 + J_1} \sqrt{(2J_0+1)(2J_1+1)} \sum_{v_2 J_2}
\begin{Bmatrix} J_2 & j & J_1 \\ I_s & j & J_0 \end{Bmatrix} \nonumber \\
 & & \times [j^{n-1} (v_2 J_2) j J_0 |\} j^n v_0 J_0]
[j^{n-1} (v_2 J_2) j J_1 |\} j^n v_1 J_1]. \label{newred}
\end{eqnarray}

This differs from the Redmond formula inasmuch as there is now a sum on the
left-hand side of the equation over $v_s$. Note that $I_s$ is fixed.
Basically, then, the sum is over all states that are present which have angular
momentum $I_s$ for the $(n+1)$-particle system. 

Of course, the fixed values of $(v_0 J_0)$ and $(v_1 J_1)$ will
lead to some restrictions on the possible values of $v_s$.

We give now an example. Consider the case $n=3$, $j=9/2$; for the 
three-particle systems, take $J_0=9/2, v_0=3$ and $J_1=11/2, v_1=3$; and 
for the angular momentum of the four-particle system, take $I_s=2$. Taking
into account that, in this case, we have two values for the seniority $v_s$,
the result of the left-hand side of Eq.~(\ref{newred}) is
$$
\begin{array}{rcr}
I_s=2, v_s=2 & \hspace{1cm} & 4\cdot(-0.128118)\cdot0.320983=-0.164495 \\
I_s=2, v_s=4 & & 4\cdot(-0.265908)\cdot0.666200=-0.708592 \\ \hline
\mbox{\rm Sum (lhs)} & & -0.873087 
\end{array}
$$

For the right-hand side of Eq.~(\ref{newred}), we obtain

$$
\begin{array}{rcr}
J_2=2 & \hspace{1cm} & -0.119633 \\
J_2=4 & & -0.789733 \\
J_2=6 & & +0.199694 \\
J_2=8 & & -0.163415 \\ \hline
\mbox{\rm Sum (rhs)} & & -0.873087
\end{array}
$$
As can be seen, we get the same result.

For the same case as above but with $v_0=1$ and $v_s=3$, we find that 
\begin{equation}
\text{lhs}=\text{rhs}=0.78674 .
\end{equation}

\section{The special case $n=2$. Application of the seniority Redmond
relation to the number of states of angular momentum $I_s$ for three identical
particles in a single $j$ shell}

For $n=2$ the two cfp's on the right-hand side of Eq.~\eqref{newred} are equal 
to 1 and $J_2=j$, i.e., the sum over $J_2$ consists of only one term. We find

\begin{eqnarray}
\lefteqn{2 \begin{Bmatrix} j & j & J_1 \\ I_s & j & J_0 \end{Bmatrix}
(-1)^{J_0+J_1} \sqrt{(2J_0+1)(2J_1+1)} =} \nonumber \\
 & & = - \delta_{J_0 J_1} \delta_{v_0 v_1} +
3 \sum_{v_s} [j^2 (v_0 J_0) j I_s |\} j^3 v_s I_s]
[j^2 (v_1 J_1) j I_s |\} j^3 v_s I_s].
\end{eqnarray}
For $J_1=J_0$ we get

\begin{eqnarray}
\frac{2}{3} \begin{Bmatrix} j & j & J_0 \\ I_s & j & J_0 \end{Bmatrix} (2J_0+1)
+ \frac{1}{3} = \sum_{v_s \alpha_s} [j^2 (v_0 J_0) j I_s |\} 
j^3 v_s I_s ]^2.
\end{eqnarray}

If we sum over $J_0$ (even) on the right-hand side, we obtain

\begin{equation}
\sum_{J_0} [j^2 (v_0 J_0) j I_s |\} j^3 v_s I_s]^2 = 1.
\end{equation}
And then, the sum over $v_s$ gives us the number of states with total angular
momentum $I_s$.

For $I_s=j$ we get a result previously obtained by Rosensteel and 
Rowe~\cite{rr03} using a quasispin formulation
\begin{equation}
\frac{1}{3} \left[ \frac{(2j+1)}{2} + 2 \sum_{J_0 \; {\rm even}}
(2J_0 + 1) \begin{Bmatrix} j & j & J_0 \\ j & j & J_0 \end{Bmatrix} \right] =
\mbox{\rm \# of states with $I_s=j$}.
\end{equation}
Ginocchio and Haxton~\cite{gh93} showed this quantity to be equal to 
$[(2j+3)/6]$, where the square brackets mean the largest integer less than what
is inside them.

For $I_s=j+1$ we get the Zhao--Arima result~\cite{za04}

\begin{equation}
\frac{1}{3} \left[ \frac{2j-1}{2} -2 \sum_{J_0\;{\rm even}} (2J_0+1)
\begin{Bmatrix} j & j & J_0 \\ j & j+1 & J_0 \end{Bmatrix} \right] =
\mbox{\rm \# of states with $I_s=j+1$},
\end{equation}
which can be shown to be $[j/3]$. The present authors have presented an 
alternative derivation of the above two results by using an $m$ 
scheme~\cite{ze05gh}. A recent preprint by Talmi also uses the $m$ scheme to go
beyond the above two examples and to prove many conjectures of Zhao and 
Arima~\cite{za03}.

\section{Isospin considerations. Application of the principal-parent Redmond
relation to problems involving neutrons and protons}

In a previous work, ``Interrelationship of isospin and angular 
momentum''~\cite{zml04}, we considered the following simple interaction in a 
single $j$ shell of neutrons and protons

\begin{equation}
\langle (j^2)^{J_A} \; V \; (j^2)^{J_A} \rangle = a \; \frac{1-(-1)^{J_A}}{2} .
\end{equation}
Since in a single $j$ shell when $J_A$ is even the isospin $T_A$ is 1, and when
$J_A$ is odd $T_A$ is 0, we see that this interaction acts only for $T_A=0$
states, i.e., only for the neutron--proton interaction in the $T_A=0$ channel.
The interaction vanishes for two neutrons or for two protons---they have 
isospin 1.

When applied to the $I=0$ states of the even--even Ti isotopes with 
configuration $[(j^2)^J_\pi (j^n)^J_\nu]^{I=0}$, the authors found the 
following expression for the interaction matrix elements:

\begin{eqnarray}
\langle [J' J']^0 \; H \;[J J]^0 \rangle /a & = & n \delta_{J J'} - n 
\sqrt{(2J+1) (2J'+1)} \nonumber \\
 & & \times \sum_{J_0} [j^{n-1} J_0 j |\} j^n J] [j^{n-1} J_0 j |\} j^n J']
\begin{Bmatrix} J_0 & j & J' \\ j & j & J \end{Bmatrix}.
\end{eqnarray}
By using the principal-parent Redmond formula [Eq.~(\ref{redmond}) of this 
work], one obtains

\begin{eqnarray}
\langle [J' J']^0 \; H \; [J J]^0 \rangle & = & (n+(-1)^{J+J'}) \delta_{J J'} 
\nonumber \\
 & & - (n+(-1)^{J+J'}) [j^n J j |\} j^{n+1} j] [j^n J' j |\} j^{n+1} j].
\end{eqnarray}
However, this can be simplified because, if we have a system of 2 protons,
then both $J$ and $J'$ must be even. 

The above result was coupled with the fact that one could also write the same
interaction in the isospin space as $a (1/4 - t(1) \cdot t(2))$. This also
vanishes for $T=1$ and is equal to a constant $a$ for $T=0$.

From the isospin point of view, it is trivial to obtain the eigenvalues for a
system of 2 protons and $n$ neutrons:

$$
\begin{array}{lcl}
\langle V \rangle = (n+1) a & & \mbox{for $T=T_{\rm min}=|N-Z|/2$,} \\
\langle V \rangle = 0 & \hspace{2cm} & \mbox{for $T=T_{\rm min}+2$.}
\end{array}
$$

The angular momentum expression~\cite{zml04} did not involve seniority. From 
what we have
seen in the previous sections, the generalization is not too difficult
\begin{eqnarray}
\langle [J' (J'v')]^0 \; H \; [J (Jv)]^0 \rangle & = & (n+1) \delta_{J J'}
\delta_{v v'} \\
 & & - (n+1) \sum_{v_f} [j^n (Jv) j |\} j^{n+1} (j v_f)]
[j^n (J' v') j |\} j^{n+1} (j v_f)]. \nonumber
\end{eqnarray}
The eigenvalue equation for this Hamiltonian is
\begin{eqnarray}
(n+1) D(J, Jv) & - & (n+1) \sum_{v_f} [j^n (Jv) j |\} j^{n+1}(j v_f)] \\
 & \times & \sum_{J'} [j^n (J'v') j |\} j^{n+1} (j v_f)] D(J',J'v') = 
\lambda D(J,Jv). \nonumber
\end{eqnarray}
However, from the isospin point of view, the eigenvalue $\lambda$ for 
$T=T_{\rm min}$ is equal to $(n+1)$. Hence, for $T=T_{\rm min}$ we obtain

\begin{equation}
\sum_{v_f} [j^n (Jv) j |\} j^{n+1} (j v_f)]
\sum_{J'} [j^n (J'v') j |\} j^{n+1} (j v_f)] D(J',J'v')=0.
\end{equation}
We can multiply by $[j^n (Jv) j |\} j^{n+1} (j v_x)]$ and sum over $v$. Thus,
using the property

\begin{equation}
\sum_v [j^n (Jv) j |\} j^{n+1} (j v_x)] [j^n (Jv) j |\} j^{n+1} (j v_f)]
= \delta_{v_f v_x},
\end{equation}
we find

\begin{equation}
\sum_{J'} [j^n (J'v') j |\} j^{n+1} (j v_x)] D^{T_{\rm min}}(J',J'v')=0
\label{orthog}
\end{equation} 
for each $v_x$ state.

What is the significance of Eq.~(\ref{orthog})? We will now show, by a 
generalization of a result of Zamick and Devi~\cite{zd99}, that this equation
expresses the fact that states with isospin $T=T_{\rm min}+2$ are orthogonal
to states with isospin $T=T_{\rm min}$.

States of 2 protons and $n$ neutrons with isospin $T_{\rm max}=T_{\rm min}+2$
are double analogs of states of ($n+2$) identical particles. This leads to the
fact that the values of the wave-function components for $T=T_{\rm max}$ are
two-particle coefficients of fractional parentage~\cite{zetal05,ze05}

\begin{equation}
D^{T_{\rm max}, I=0, v_f} (J, Jv)=[j^n(Jv)j^2(J)|\} j^{n+2} I=0 \; v_f].
\end{equation}
For a system of $n+2$ identical particles (neutrons), we can write

\begin{equation}
|j^{n+2} \rangle^{I\; v_f} = \sum_{J_0\; v_0\; v_1}
[j^n(J_0 v_0) j^2(J_0 v_1)|\} j^{n+2} I v_f] \;
|(j^n)^{J_0 v_0} (j^2)^{J_0 v_1} \rangle^{I v_f}.
\end{equation}

We can, however, reach this result in two stages with successive one-particle
cfp's:

\begin{eqnarray}
|j^{n+2} \rangle^{I \; v_f} = \sum_{J_0 v_0} \sum_{v_3} & &
[j^{n+1}(j v_3) j |\} j^{n+2} I v_f] \;
[j^n(J_0 v_0) j |\} j^{n+1} j v_3] \nonumber \\
 & & \times U(J_0 j I j;j J_0) \; |(j^n)^{J_0 v_0} (j^2)^{J_0} \rangle^{I v_f}.
\end{eqnarray}
In the above, $U$ is a unitary Racah coefficient. For $I=0$ the value of $U$ 
is 1.

So far we have
\begin{equation}
[j^n(J_0 v_0) j^2(J_0 v_1) |\} j^{n+2} I=0 \; v_f] = \sum_{v_3}
[j^n(J_0 v_0) j |\} j^{n+1}(j v_3)] \; 
[j^{n+1}(j v_3) j |\} j^{n+2} I=0 \; v_f].
\end{equation}
However, it can be shown, e.g. in Bayman and Lande~\cite{bl66}, that the $(n+1)
\rightarrow (n+2)$ cfp above to a final state with $I=0$ is 1 for $v_3=v_f-1$
and 0 otherwise, except when $v_f=0$, in which case $v_3=1$. So we have
\begin{eqnarray}
[j^n(J_0 v_0) j^2(J_0 v_1) |\} j^{n+2} I=0 \; v_f] & = & 
[j^n(J_0 v_0) j |\} j^{n+1} (j,v_f-1)] \ , \hspace{0.7cm} 
\mbox{for $v_f > 0$}, \\
 & = & [j^n(J_0 v_0) j |\} j^{n+1} (j,v=1)] ~, \hspace{0.8cm} 
\mbox{for $v_f=0$}.
\end{eqnarray}

Thus, in Eq.~(\ref{orthog}) we can replace the one-particle cfp $n
\rightarrow (n+1)$ by the two-particle cfp $n \rightarrow (n+2)$. The latter
is more obviously identified with the wave function of a state with $I=0$ and
$T=T_{\rm max}$.

\section{Closing remarks}

We have here discussed both principal-parent coefficients of fractional
parentage and those obtained by seniority schemes. The former are easier to
calculate, but form an overcomplete set; while the latter form a complete
orthonormal set. The original Redmond formulation gives $n \to n+1$ cfp's 
which are obtained by a principal-parent classification via the $n-1 \to n$
cfp's with no clear classification. Our main result in this work was to obtain
Redmond-type relations (see Eqs.~\eqref{pp2scfp} and \eqref{newred}) in which 
we have seniority cfp's on both sides of the equation. A new feature is that on
the left-hand side of Eq.~\eqref{newred} we have, for fixed final angular 
momentum, a sum over all possible final seniorities.

We have noted that both principal parent and seniority cfp's have their uses.
For the former, we noted earlier works which showed that, by constructing
cfp's to non-existent states, e.g., $(f_{7/2}^3),J=13/2$, we obtain 
conditions on 6$j$ symbols. In turn, the vanishing of these cfp's was put to 
use in a completely different problem by Robinson and Zamick~\cite{rz}, namely
to explain a partial dynamical symmetry for a system of two neutrons and one 
proton (likewise two protons and one neutron) when the $T=0$ two-body 
interaction is set to zero and only $T=1$ two-body matrix elements are used.

We applied our new Redmond relation to the problem of the number of states of a
given angular momentum in a $(j^3)$ configuration. Previously, we had results 
only up to $j\leq 7/2$, but with the new Redmond relation we have it for all
$j$. We also used this relation to generalize a relation by Zamick, Mekjian and
Lee~\cite{zml04}, again for $j \leq 7/2$, to higher $j$ values, where states of
a given
angular momentum occur more than once for a three-particle system. We thereby 
obtain conditions on the wave functions of states of mixed neutrons and protons
which boil down to the fact that states of higher isospin are orthogonal to
states of lower isospin.

While our new Redmond relation at first sight appears more complicated than the
original one, because of the sum over final seniorities on the left-hand side,
we find that this sum can be used to obtain closure and ultimately can lead to
simple results.

\appendix

\section{}
\label{app}

We here offer a proof of Eq.~(\ref{pp2scfp}). We will do it for a $(g^3_{9/2})$
three-particle system. The proof for any number of particles and other 
configurations is essentially the same.

The wave function for the three-particle system with a principal parent $[J_0]$
for the two particles is [see Eq.~(\ref{antiwf})]
\begin{equation}
\Psi^J [J_0] = N [J_0] (1-P_{12}-P_{13}) \left[ [12]^{J_0} 3 \right]^J ,
\end{equation}
where $J$ is the total angular momentum of the three particles. These $\Psi^J
[J_0]$'s are an overcomplete set; e.g., for $J=9/2$ there are five $\Psi^J 
[J_0]$'s, but only two independent wave functions. Althought not necessary, we
can separate the two into states with definite seniority $v=1$ and $v=3$.

The following relation must hold between the principal parent wave functions
and the seniority wave functions
\begin{equation}
\Psi^J [J_0] = C[J_0] \Psi^J (v=1) + D[J_0] \Psi^J (v=3) ,
\end{equation}
or in more detail
\begin{subequations}
\begin{eqnarray}
\lefteqn{\sum_{J_1} [j^2 (J_1) j J |\} j^3 [J_0] J] \left[ [12]^{J_1} 
 3 \right]^J = } \\
 & & \sum_{J_1} \left\{ C[J_0] [j^2 (J_1) j J |\} j^3 v=1,J] \left[ 
 [12]^{J_1} 3 \right]^J \right. \\ 
 & & \left. \text{ } + D[J_0] [j^2 (J_1) j J |\} j^3 v=3,J] \left[ 
 [12]^{J_1} 3 \right]^J \right\} .
\end{eqnarray}
\end{subequations}
This leads to the following relation between cfp's
\begin{equation}
[j^2 (J_1) j J |\} j^3 [J_0] J] = C[J_0] [j^2 (J_1) j J |\} j^3 v=1,J ] +
D[J_0] [j^2 (J_1) j J |\} j^3 v=3,J],
\end{equation}
with $C$ and $D$ independent of $J_1$.

By taking overlaps, we see
\begin{equation}
C [J_0] = N [J_0] \langle \Psi^J (v=1) | (1-P_{12}-P_{13} \left[ [12]^{J_0}
3 \right]^J \rangle .
\end{equation}
Since $\Psi (v=1)$ is totally antisymmetric, this leads to
\begin{eqnarray}
C [J_0] & = & 3 N [J_0] \langle \Psi (v=1) | \left[ [12]^{J_0} 3 \right]^J 
\rangle \nonumber \\
 & = & 3 N [J_0] [j^2 (J_0) j J |\} j^3 v=1,J] .
\end{eqnarray}
Likewise
\begin{equation}
D [J_0] = 3 N [J_0] [j^2 (J_0) j J |\} j^3 v=3,J].
\end{equation}
Thus, we have
\begin{equation}
[j^2 (J_1) j J |\} j^3 [J_0] J] = 3 N [J_0] \sum_{v=1,3} [j^2 (J_0) j J |\}
j^3 v J] [j^2 (J_1) j J |\} j^3 v J ].
\end{equation}

But, from Eq.~(\ref{cfpnf}), we see that
\begin{equation}
3 N [J_0] = \frac{1}{[j^2 (J_0) j J |\} j^3 [J_0] J]}.
\end{equation}
By cross multiplication, we get the result we are after---Eq.~(\ref{pp2scfp}). 

Once this has been shown, the Redmond relation of Eq.~(\ref{newred}) follows
because, as discussed in the text after Eq.~(\ref{pp2scfp}), the cfp's on the 
right-hand side of Eq.~(\ref{newred}) for the $n$-particle system (from $n-1$
to $n$) have been constructed with definite seniority.

\begin{acknowledgments}
We thank Igal Talmi for his explanation of the Redmond formula and Ben Bayman
for his comments. This work was 
supported by the U.S. Dept.~of~Energy under Grant No.~DE-FG0105ER05-02. A.E. is
supported by a grant financed by the Secretar\'{\i}a de Estado de Educaci\'on y
Universidades (Spain) and cofinanced by the European Social Fund.
\end{acknowledgments}


\begin{thebibliography}{99}
\bibitem{r54} P.~J.~Redmond, Proc. R. Soc. London A {\bf 222}, 84 (1954).

\bibitem{st63} A. de~Shalit and I.~Talmi, {\it Nuclear Shell Theory},
Academic Press, New York (1963).

\bibitem{t93} I.~Talmi, {\it Simple Models of Complex Nuclei}, Harwood 
Academics Press, Switzerland (1993).

\bibitem{bl66} B. F. Bayman and A.~Lande, Nucl. Phys. {\bf 77}, 1 (1966).

\bibitem{r43} G.~Racah, Phys. Rev. {\bf 63}, 367 (1943).

\bibitem{r49} G.~Racah, Phys. Rev. {\bf 76}, 1352 (1949).

\bibitem{l80} R. D. Lawson, {\it Theory of the Nuclear Shell Model}, Clarendon
Press, Oxford (1980).

\bibitem{rz} S. J. Q. Robinson and L. Zamick, Phys. Rev. C {\bf 63}, 064316
(2001); S.~J.~Q.~Robinson and L.~Zamick, Phys. Rev. C {\bf 64}, 057302 (2001).

\bibitem{rr03} G.~Rosensteel and D.J.~Rowe, Phys. Rev. C {\bf 67}, 014303 
(2003).

\bibitem{gh93} J.N.~Ginocchio and W.C.~Haxton, {\it Symmetries in Science VI}, 
ed. by B.~Gruber and M.~Ramek, Plenum, New York (1993) .

\bibitem{za04} Y. M.~Zhao and A.~Arima, Phys. Rev. C {\bf 70}, 034306 (2004).

\bibitem{ze05gh} L. Zamick and A. Escuderos, Phys. Rev. C {\bf 71}, 054308
(2005).

\bibitem{za03} Y. M. Zhao and A. Arima, Phys. Rev. C {\bf 68}, 044310 (2003).

\bibitem{zml04} L.~Zamick, A.~Z.~Mekjian, and S.~J.~Lee, J. Korean Phys. Soc.
{\bf 47}, 18 (2005).

\bibitem{zd99} L.~Zamick and Y.~D.~Devi, Phys. Rev. C {\bf 60}, 054317 (1999).

\bibitem{zetal05} L. Zamick, A.~Escuderos, S.~J.~Lee, A.~Z.~Mekjian, E.~Moya 
de~Guerra, A.~A.~Raduta, and P.~Sarriguren, Phys. Rev. C {\bf 71}, 034317 
(2005).

\bibitem{ze05} L. Zamick and A.~Escuderos, Phys. Rev. C {\bf 71}, 014315 
(2005).

\end{thebibliography}
\end{document}